\documentclass[aps,prb,twocolumn,groupedaddress,showpacs]{revtex4}
\usepackage{graphicx}
\usepackage{dcolumn}
\usepackage{bm}

\begin{document}

\title{Spin-orbital gap of multiorbital antiferromagnet}

\author{Hiroaki Onishi and Takashi Hotta}

\affiliation{%
Advanced Science Research Center,
Japan Atomic Energy Agency,
Tokai, Ibaraki 319-1195, Japan}

\date{July 24, 2006}

\begin{abstract}
In order to discuss the spin-gap formation in a multiorbital system,
we analyze an $e_{\rm g}$-orbital Hubbard model
on a geometrically frustrated zigzag chain
by using a density-matrix renormalization group method.
Due to the appearance of a ferro-orbital arrangement,
the system is regarded as a one-orbital system,
while the degree of spin frustration is controlled by
the spatial anisotropy of the orbital.
In the region of strong spin frustration, we observe
a finite energy gap between ground and first-excited states,
which should be called a {\it spin-orbital} gap.
The physical meaning is clarified by an effective Heisenberg spin model
including correctly the effect of the orbital arrangement
influenced by the spin excitation.
\end{abstract}



\maketitle

\section{Introduction}

Exotic properties of low-dimensional antiferromagnets have
continued to attract much attention
in the research field of condensed matter physics,
since strong quantum spin fluctuation destroys classical N\'eel ordering
even at zero temperature, while it leads to various types of
quantum disordered states.~\cite{QSS04}
In fact, substantial advances in material synthesis and experimental probes
have made it possible to gain deep insights into
strange spin-liquid phases and quantum phase transitions
in realistic low-dimensional compounds.
In particular, a class of spin-gapped antiferromagnets has
provided rich macroscopic quantum phenomena such as
spin-Peierls transition and impurity-induced magnetic ordering
in CuGeO$_{3}$,~\cite{Hase1993,Masuda2000,Fukuyama1996,Onishi2000}
and Bose-Einstein condensation of magnons
in TlCuCl$_{3}$.~\cite{Oosawa1999,Nikuni2000,Ruegg2003}

Microscopic aspects of the spin-gapped phase
in low-dimensional antiferromagnets have been extensively discussed
on the basis of the antiferromagnetic (AFM) Heisenberg model.
In general, a spin gap is defined as
the energy difference between singlet ground and
triplet first-excited states.
The existence of the finite spin gap can be understood
by the concept of a valence-bond state.
One typical example is a spin-1/2 Heisenberg zigzag
chain.~\cite{MG1969,Haldane1982,Tonegawa1987,Okamoto1992,White1996}
In fact, the combined effects of quantum fluctuation
and geometrical frustration yield a transition
from a gapless spin-liquid phase to a gapped dimer phase.
In the gapped dimer phase,
a valence bond of neighboring spins is stabilized,
but the correlations among the valence bonds are significantly
suppressed due to the effect of spin frustration,
implying that finite energy is required to excite the valence bond
from singlet to triplet.

In experiments,
as for a model material of the zigzag spin chain,
several compounds have been reported,
such as SrCuO$_2$,~\cite{Matsuda1995,Matsuda1997}
Cu(ampy)Br$_2$,~\cite{Kikuchi2000}
and (N$_2$H$_5$)CuCl$_3$.~\cite{Hagiwara2001,Maeshima2003}
However, a candidate material with a significant spin gap
has not been found,
since the frustration effect is much weak in these materials.
In such a circumstance,
in order to realize the spin-gapped dimer phase
in a region of the strong spin frustration,
it is important to explore the way
to control the degree of the spin frustration.
In this context, it is expected that
orbital degree of freedom is a key ingredient
to cause highly non-uniform spin exchange interactions. 
Indeed, since $d$- and $f$-electron orbitals in recently focused materials
are spatially anisotropic,
the sign and magnitude of spin exchange interactions
depend on the relative positions of ions and occupied orbitals.
In such a case, we envision that
orbital ordering occurs so as to minimize the magnetic energy of
the spin system.
In fact, we have found that orbital ordering actually relaxes
the frustration effect in the exchange interaction due to
the spatial anisotropy of relevant orbital, even though
the lattice originally exhibits
a geometrically frustrated structure.~\cite{Onishi2005a,Onishi2005b,Onishi2006}

After the occurrence of orbital ordering,
one may naively expect that
the system is described by an effective spin system
on the frozen background of orbital arrangement.
It is true that we can intuitively understand
what kind of magnetic correlation develops
in the static orbital ordering.
However, since spin and orbital degrees of freedom strongly
interact with each other at the same energy scale,
spin dynamics is influenced by orbital one in principle,
unless orbital degree of freedom is completely quenched.
Such interplay of spin and orbital dynamics has been discussed
for spin-orbital models.~\cite{Brink1998,Bala2001}
Here, we point out that even though we observe a spin gap,
this would not be a pure spin excitation
but includes the changes of both spin and orbital states.
The spin gap is definitely given by the energy difference
between singlet ground and triplet first-excited
states, but the orbital arrangement in the ground and excited
states can be different.
We believe that it is an intriguing issue to clarify
how the spin excitation is described in such a
variable background of orbital arrangement.

In this paper, we investigate the spin gap of
an $e_{\rm g}$-orbital Hubbard model
on a zigzag chain by numerical techniques.
Due to the appearance of a ferro-orbital (FO) arrangement,
the system can be always regarded as a one-orbital system
irrespective of the level splitting.
When two orbitals are degenerate,
the zigzag chain is decoupled to a double-chain spin system
to suppress the spin frustration due to the orbital anisotropy.
On the other hand,
the orbital anisotropy is controlled by the level splitting,
leading to the revival of the spin frustration
and a finite spin-excitation gap.
However, the orbital arrangement is found to change
according to the spin excitation,
suggesting an important concept of
dynamical interplay between spin and orbital degrees of freedom.

The organization of this paper is as follows.
In Sec.~II,
we describe the model Hamiltonian and numerical techniques.
In Sec.~III,
we show the results of the spin and orbital structures
in the ground state.
Taking account of the orbital arrangement,
we will introduce an effective Heisenberg model.
Then, in Sec.~IV,
we discuss the spin-excited states
based on the effective model.
Finally, the paper is summarized in Sec.~V.

\section{Model and Method}

\begin{figure}[t]
\includegraphics[width=0.8\linewidth]{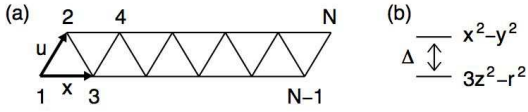}
\caption{
(a) Lattice arrangement and site numbering of $N$-site zigzag chain.
(b) Schematic view of level splitting between $e_{\rm g}$ orbitals.
}
\end{figure}

Let us consider $e_{\rm g}$ orbitals on a zigzag chain with $N$ sites.
The lattice is composed of equilateral triangles
and located in the $xy$ plane [see Fig.~1(a)].
Note that the zigzag chain is considered as a two-chain system
connected by a zigzag path.
We consider the case of one electron per site (quarter filling),
which is corresponding to, for example, low-spin state of Ni$^{3+}$ ion
with $3d^7$ configuration.

When the Hund's rule interaction is small,
the ground state is found to be paramagnetic,~\cite{Onishi2005a,Onishi2005b}
which is relevant to a geometrically frustrated antiferromagnet.
Here, we study the property of the paramagnetic phase
and simply ignore the Hund's rule interaction,
but the results do not change qualitatively for quarter filling case.
Then, the Hamiltonian is given by
\begin{eqnarray}
\label{eq:ham}
 H
 &=&
 \sum_{{\bf i},{\bf a},\tau,\tau',\sigma}
 t_{\tau\tau'}^{\bf a}
 d_{{\bf i}\tau\sigma}^{\dag} d_{{\bf i}+{\bf a}\tau'\sigma}
 -(\Delta/2) \sum_{\bf i}
 (\rho_{{\bf i}a}-\rho_{{\bf i}b})
 \nonumber\\
 &&
 +U \sum_{{\bf i},\tau}
 \rho_{{\bf i}\tau\uparrow} \rho_{{\bf i}\tau\downarrow}
 +U' \sum_{{\bf i}} 
 \rho_{{\bf i}a} \rho_{{\bf i}b},
\end{eqnarray}
where $d_{{\bf i}a\sigma}$ ($d_{{\bf i}b\sigma}$)
is the annihilation operator for an electron with spin $\sigma$
in the 3$z^2$$-$$r^2$ ($x^2$$-$$y^2$) orbital at site ${\bf i}$,
$\rho_{{\bf i}\tau\sigma}$=%
$d_{{\bf i}\tau\sigma}^{\dag}d_{{\bf i}\tau\sigma}$,
$\rho_{{\bf i}\tau}$=%
$\sum_{\sigma}\rho_{{\bf i}\tau\sigma}$,
${\bf a}$ is the vector connecting adjacent sites,
and $t_{\tau\tau'}^{\bf a}$ is the hopping amplitude
between $\tau$ and $\tau'$ orbitals
along the ${\bf a}$ direction.

The hopping amplitudes are evaluated from the overlap integral
between $e_{\rm g}$ orbitals in adjacent sites,
\cite{Slater1954,Hotta,Dagotto2001}
which are given by
\begin{equation}
 t_{aa}^{\bf x}=t/4, \
 t_{ab}^{\bf x}=t_{ba}^{\bf x}=-\sqrt{3}t/4, \
 t_{bb}^{\bf x}=3t/4
\end{equation}
for the ${\bf x}$ direction,
\begin{equation}
 t_{aa}^{\bf u}=t/4, \
 t_{ab}^{\bf u}=t_{ba}^{\bf u}=\sqrt{3}t/8, \
 t_{bb}^{\bf u}=3t/16
\end{equation}
for the ${\bf u}$ direction,
and $t_{\tau\tau'}^{{\bf u}-{\bf x}}$=$t_{\tau\tau'}^{\bf u}$.
Thus, the hopping amplitudes depend on
the bond direction and occupied orbitals.
Hereafter, $t$ is taken as the energy unit.

The second term of $H$ denotes the crystalline electric field (CEF)
potentials. 
Although the CEF potential should be evaluated by the sum of
electrostatic potentials from ligand ions surrounding the
transition metal ion,\cite{Hotta}
here we consider the level splitting $\Delta$
between 3$z^2$$-$$r^2$ and $x^2$$-$$y^2$ orbitals
due to the tetragonal CEF effect,
as shown in Fig.~1(b).
In the third and fourth terms,
$U$ and $U'$ indicate intraorbital and interorbital
Coulomb interactions, respectively.
For the local interaction parameters,
the relation $U$=$U'$ holds due to the rotational invariance
in the local orbital space,
when we ignore the Hund's rule coupling and pair hopping
parameters.~\cite{Dagotto2001}

We analyze the model (\ref{eq:ham}) by exploiting
a finite-system density-matrix renormalization group (DMRG) method
with open boundary conditions.~\cite{White1992,Schllwock2005}
Note that due to the two orbitals in one site,
the number of bases for the single site is 16
and the size of the superblock Hilbert space grows as $m^2$$\times$$16^2$,
where $m$ is the number of states kept for each block.
Here, we treat each orbital as a site
to reduce the size of the superblock Hilbert space to $m^2$$\times$$4^2$.
Then, the original zigzag chain is considered as a four-leg ladder
and the finite-system sweep is performed along a one-dimensional path,
as shown in Fig.~2.
We note that in the superblock configuration,
system and environment blocks are connected
due to the on-site Coulomb interaction
as well as the inter-site electron hopping.
Thus, we can increase $m$ and accelerate the calculations
due to the reduction of matrix size.
In the present calculations,
we keep up to $m$=300 states
and the truncation error is kept around $10^{-6}$ or smaller.

\begin{figure}[t]
\includegraphics[width=0.7\linewidth]{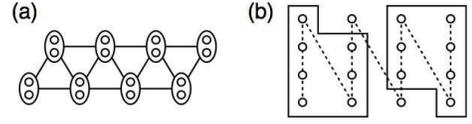}
\caption{
(a) Original zigzag chain including two orbitals at each site.
(b) Superblock configuration, in which each orbital is regarded as a site.
The dotted line denotes the path for the finite-system algorithm.
}
\end{figure}

In order to determine uniquely the orbital state,
it is useful to introduce an angle $\theta_{\bf i}$
to characterize the orbital shape at each site.
We define new operators for electrons in the rotated frame as
\cite{Hotta1998,Kugel1972,Kugel1973,complex-orbital}
\begin{equation}
 \label{eq:new-op}
  \left(
   \begin{array}{l}
    \tilde{d}_{{\bf i}\alpha\sigma} \\
    \tilde{d}_{{\bf i}\beta\sigma}
   \end{array}
  \right)
  =
  R(\theta_{\bf i})
  \left(
   \begin{array}{l}
    d_{{\bf i}a\sigma} \\
    d_{{\bf i}b\sigma}
   \end{array}
  \right),
\end{equation}
where the transformation matrix $R(\theta_{\bf i})$ is given by
\begin{equation}
 R(\theta_{\bf i})
 =
 e^{i\theta_{\bf i}/2}
 \left(
  \begin{array}{ll}
  \cos(\theta_{\bf i}/2) & \sin(\theta_{\bf i}/2) \\
  -\sin(\theta_{\bf i}/2) & \cos(\theta_{\bf i}/2)
  \end{array}
 \right).
\end{equation}
For reference, the correspondence of the angle $\theta_{\bf i}$
to the orbitals is summarized in Table~I.

Now the problem is how to determine the set of angles $\{\theta_{\bf i}\}$.
Naively thinking, a straightforward way is to find the set of
$\{\theta_{\bf i}\}$ which minimizes the energy of the target state.
However, due to the rotational symmetry in the orbital space,
the energy does $not$ depend on $\{\theta_{\bf i}\}$
in the electronic model including only Coulomb interactions.\cite{Hotta2000}
In such a situation, in order to determine $\{\theta_{\bf i}\}$,
it is necessary to maximize the orbital correlations,
as shown in Ref.~\onlinecite{Hotta2000}.

In general, to grasp what types of correlations develop,
we investigate appropriate correlation functions.
As for the orbital state, in the pseudo-spin representation for two orbitals,
it is necessary to measure the orbital structure factor, defined by
\begin{equation}
 T({\bf q})=
 \sum_{{\bf i},{\bf j}}
 \langle \tilde{T}_{\bf i}^{z}\tilde{T}_{\bf j}^{z} \rangle
 e^{i{\bf q}\cdot({\bf i}-{\bf j})}/N
\end{equation}
with
$\tilde{T}_{\bf i}^z$=$\sum_{\sigma}
(\tilde{d}_{{\bf i}\alpha\sigma}^{\dag}
 \tilde{d}_{{\bf i}\alpha\sigma}
 $$-$$
 \tilde{d}_{{\bf i}\beta\sigma}^{\dag}
 \tilde{d}_{{\bf i}\beta\sigma})/2$,
where $\langle \cdots \rangle$ denotes the expectation value.
For each ${\bf q}$, we find the set of $\{\theta_{\bf i}\}$
to maximize $T({\bf q})$.
Among them, the optimal set of $\{\theta_{\bf i}\}$ should be
given by the one with the largest value of $T({\bf q})$.
Namely, the optimal orbital at each site is determined
so as to maximize the amount of the orbital correlation.
We note that the evaluation of $T({\bf q})$
for any set of $\{\theta_{\bf i}\}$ can be carried out
with just a single DMRG calculation
by keeping appropriately the components of orbital correlations
in terms of the original orbital basis set,
due to the relation
$\tilde{T}_{\bf i}^z=
 T_{\bf i}^z \cos\theta_{\bf i}+T_{\bf i}^x \sin\theta_{\bf i}$
with
$T_{\bf i}^z=\sum_{\sigma}
 (d_{{\bf i}a\sigma}^{\dag}d_{{\bf i}a\sigma}
 -d_{{\bf i}b\sigma}^{\dag}d_{{\bf i}b\sigma})/2$
and
$T_{\bf i}^x=\sum_{\sigma}
 (d_{{\bf i}a\sigma}^{\dag}d_{{\bf i}b\sigma}
 +d_{{\bf i}b\sigma}^{\dag}d_{{\bf i}a\sigma})/2$.

Although the set of $\{\theta_{\bf i}\}$ cannot be determined from
the energetical discussion of the two-orbital model,
it is useful to consider the energy of the effective one-orbital model
composed of the lower-energy orbitals specified by $\{\theta_{\bf i}\}$.
It has been found that the energy of the effective one-orbital model
is actually minimized when the set of $\{\theta_{\bf i}\}$ is
determined so as to maximize the orbital correlation.
Thus, we believe that our scheme to maximize the orbital correlation
works well to find the optimal orbital state.

\begin{table}[t]
\caption{
Angle $\theta_{\bf i}$ and the corresponding orbitals.
}
\begin{ruledtabular}
\begin{tabular}{ccc}
$\theta_{\bf i}$ & $\alpha$ orbital & $\beta$ orbital \\
\hline
0        & 3$z^2$$-$$r^2$ & $x^2$$-$$y^2$  \\
$\pi$/3  & $y^2$$-$$z^2$  & 3$x^2$$-$$r^2$ \\
2$\pi$/3 & 3$y^2$$-$$r^2$ & $z^2$$-$$x^2$  \\
$\pi$    & $x^2$$-$$y^2$  & 3$z^2$$-$$r^2$ \\
4$\pi$/3 & 3$x^2$$-$$r^2$ & $y^2$$-$$z^2$  \\
5$\pi$/3 & $z^2$$-$$x^2$  & 3$y^2$$-$$r^2$ \\
\end{tabular}
\end{ruledtabular}
\end{table}

\section{Ground State}

\begin{figure}[t]
\includegraphics[width=1.0\linewidth]{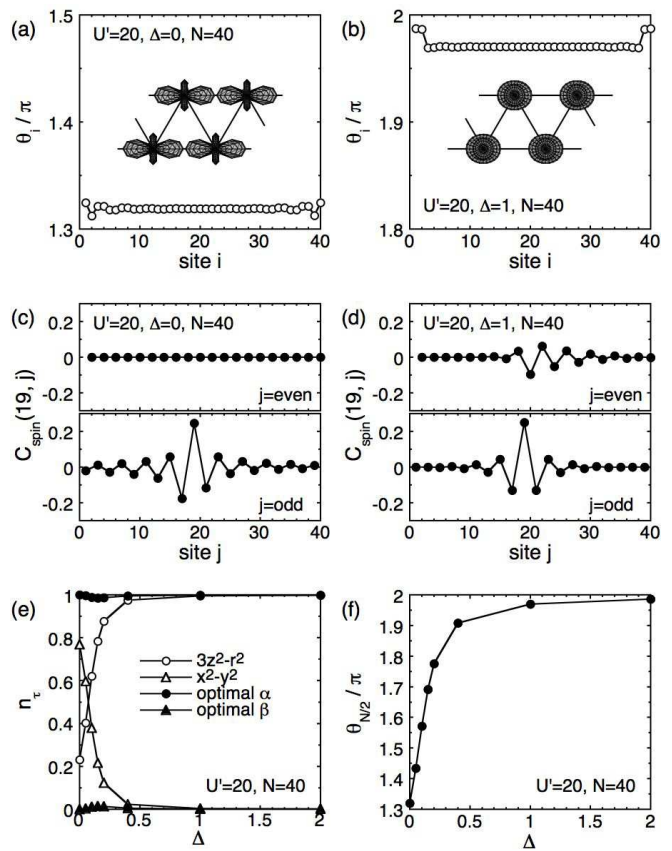}
\caption{
DMRG results for orbital states and spin correlations
in the ground state.
The site dependence of $\{\theta_i\}$
representing orbital arrangement for
(a) $\Delta$=0 and (b) $\Delta$=1.
Spin correlation function
measured from the center site of the lower chain for
(c) $\Delta$=0 and (d) $\Delta$=1.
(e) Electron densities in $3z^2$$-$$r^2$, $x^2$$-$$y^2$,
and optimal orbitals as a function of $\Delta$.
(f) Angle $\theta_{\bf i}$ at the center site of the system
as a function of $\Delta$.
}
\end{figure}

\subsection{Spin and orbital structures}

Let us first clarify the orbital structure
in the spin-singlet ground state,
an important issue to understand a key role of
orbital anisotropy in geometrically frustrated systems.
Irrespective of $\Delta$, we observe that $T({\bf q})$ becomes maximum
at ${\bf q}$=0, indicating a FO structure.
In Figs.~3(a) and 3(b), we show the optimal set of $\{\theta_{\bf i}\}$
for $\Delta$=0 and 1, respectively.
The corresponding orbital arrangement
in the bulk is also depicted.
At $\Delta$=0, two orbitals are degenerate, but in general,
orbital degeneracy is lifted to lower the energy by electron hopping,
leading to an optimal path for the electron motion.
Indeed, there appears a FO structure characterized by
$\theta_{\bf i}$/$\pi$$\sim$1.32,
indicating that 3$x^2$$-$$r^2$ orbital is selectively occupied.
The shape of the occupied orbital extends
just along the {\bf x} direction, not along the zigzag path,
leading to a double-chain structure for the electron motion.
Then, the spin exchange interaction along the zigzag path
is expected to be suppressed in comparison with
that along the double-chain direction.
To clarify this point,
we investigate the spin correlation function
\begin{equation}
 C_{\rm spin}({\bf i},{\bf j})=
 \langle S_{\bf i}^z S_{\bf j}^z \rangle
\end{equation}
with
$S_{\bf i}^z$=$\sum_{\tau}
(\rho_{{\bf i}\tau\uparrow}$$-$$\rho_{{\bf i}\tau\downarrow})/2$.
In Fig.~3(c),
the correlations measured from the center site
of the lower chain is shown for $\Delta$=0.
We find that there exists an AFM correlation
between intrachain sites in each of the double chain,
while the spin correlation between the two chains is much weak.
Thus, the zigzag chain is decoupled to a double-chain system
due to the orbital anisotropy in terms of the spin correlation.

On the other hand,
the orbital anisotropy almost disappears at $\Delta$=1,
since electrons favorably occupy the lower 3$z^2$$-$$r^2$ orbital,
which is isotropic in the $xy$ plane, as shown in Fig.~3(b).
In accordance with the variation of the orbital shape,
electrons turn to move along the zigzag path
as well as the double-chain direction.
Then, the spin correlation appears between the two chains,
as shown in Fig.~3(d).
Note, however, that orbital degree of freedom is not perfectly
quenched due to the level splitting.
In fact, it is found that optimal orbitals at the edges
deviate from those in the bulk.
Namely, the orbital shape in the bulk extends to
the double-chain direction in some degree,
while it changes to become rather isotropic around the edges
so that electrons can smoothly move near the edges.
Thus, orbital degree of freedom is still effective, and
there occurs the adjustment of the optimal orbital
to the environmental lattice inhomogeneity.

In order to gain insight into the change in the orbital state
due to the level splitting,
we investigate the electron density in each orbital
\begin{equation}
 n_{\tau}=\sum_{\bf i}\langle \rho_{{\bf i}\tau} \rangle/N,
\end{equation}
as shown in Fig.~3(e).
As $\Delta$ is increased,
electrons are forced to be accommodated
in the lower 3$z^2$$-$$r^2$ orbital,
but the electron density
in each of 3$z^2$$-$$r^2$ and $x^2$$-$$y^2$ orbitals
is found to change gradually without any anomalous behavior.
Accordingly, the optimal $\alpha$ orbital state is smoothly transformed
from 3$x^2$$-$$r^2$ to 3$z^2$$-$$r^2$ with increasing $\Delta$,
as shown in Fig.~3(f).
On the other hand, we observe that
optimal $\alpha$ orbitals are occupied,
while $\beta$ orbitals are almost vacant,
irrespective of $\Delta$.
This behavior is well understood by regarding the level splitting
as the fictitious magnetic field in the orbital space.
Namely, the direction of fully polarized orbital moment is
rotated by the magnetic field.
Thus, the present system is always regarded as a one-orbital system
composed of optimal $\alpha$ orbitals,
although we have considered the multi-orbital system.

\subsection{Effective spin model}

Let us here consider the effect of the appearance of
an orbital arrangement on magnetic properties.
For finite $\Delta$,
the lower 3$z^2$$-$$r^2$ orbital is occupied in the atomic limit,
but strong electron-electron correlations cause
a mixed orbital state of 3$z^2$$-$$r^2$ and $x^2$$-$$y^2$ orbitals.
In fact, we have found that
one optimal orbital becomes relevant due to the FO arrangement,
while the relevant orbital is controlled by $\Delta$.
In order to understand magnetic properties of such a one-orbital system,
it is useful to consider an effective model in the strong-coupling limit.
For this purpose,
we first express the Hamiltonian
in terms of new operators by substituting Eq.~(\ref{eq:new-op}) as
\begin{eqnarray}
\label{eq:ham-new-op}
 {\tilde H}
 &=&
 \sum_{{\bf i},{\bf a},\tau,\tau',\sigma}
 \tilde{t}_{\tau\tau'}^{{\bf i},{\bf i}+{\bf a}}
 \tilde{d}_{{\bf i}\tau\sigma}^{\dag}
 \tilde{d}_{{\bf i}+{\bf a}\tau'\sigma}
 +U \sum_{{\bf i},\tau}
 \tilde{\rho}_{{\bf i}\tau\uparrow}\tilde{\rho}_{{\bf i}\tau\downarrow}
 \nonumber\\
 &&
 +U' \sum_{{\bf i}}
 \tilde{\rho}_{{\bf i}\alpha}\tilde{\rho}_{{\bf i}\beta}
 - \Delta \sum_{\bf i}
 (\cos\theta_{\bf i}{\tilde T}_{\bf i}^z
 + \sin\theta_{\bf i}{\tilde T}_{\bf i}^x),
\end{eqnarray}
where
$\tilde{\rho}_{{\bf i}\tau\sigma}$=$
{\tilde d}_{{\bf i}\tau\sigma}^{\dag}{\tilde d}_{{\bf i}\tau\sigma}$,
${\tilde \rho}_{{\bf i}\tau}$=$
\sum_{\sigma}{\tilde \rho}_{{\bf i}\tau\sigma}$,
${\tilde T}_{\bf i}^x=
\sum_{\sigma}
(\tilde{d}_{{\bf i}\alpha\sigma}^{\dag}\tilde{d}_{{\bf i}\beta\sigma}
+\tilde{d}_{{\bf i}\beta\sigma}^{\dag}\tilde{d}_{{\bf i}\alpha\sigma})/2$,
and the summation for the orbital index is taken
over optimal $\alpha$ and $\beta$ orbitals in the rotated frame.
The hopping amplitude is given by
\begin{equation}
 \tilde{t}_{\tau\tau'}^{{\bf i},{\bf i}+{\bf a}}=
 \sum_{\mu,\nu}
 [R(\theta_{{\bf i}})]_{\tau\mu}
 t_{\mu\nu}^{\bf a}
 [R^{-1}(\theta_{{\bf i}+{\bf a}})]_{\nu\tau'}.
\end{equation}
We note that $U$ and $U'$ ($U$=$U'$) in Eq.~(\ref{eq:ham-new-op}) are
given by the same ones in the original Hamiltonian Eq.~(\ref{eq:ham}),
since the orbital basis set is transformed
within the rotational invariant orbital space,
and thus, the Coulomb integrals
are expressed in the same form by Racah parameters.~\cite{Dagotto2001}
Note also that,
as mentioned in the previous subsection,
the level splitting is regarded as the fictitious magnetic field
for the orbital moment $\tilde{T}_{\bf i}^z$,
while there appears the transverse field for $\tilde{T}_{\bf i}^x$.

Magnetic and orbital ordering phenomena
in multi-orbital systems have been actively discussed
on the basis of an effective spin-orbital model,
the so-called Kugel-Khomskii
model.~\cite{Kugel1972,Kugel1973}
Here, we intend to obtain an understanding
of magnetic properties on the orbital-arranged background
composed of the relevant $\alpha$ orbital.
Thus, we consider a simplified spin model,
by ignoring the irrelevant $\beta$ orbital.
Then, we keep only the $\alpha$-orbital parts
in Eq.~(\ref{eq:ham-new-op}), such as
\begin{eqnarray}
 {\tilde H}_{\alpha}
 &=&
 \sum_{{\bf i},{\bf a},\sigma}
 \tilde{t}_{\alpha\alpha}^{{\bf i},{\bf i}+{\bf a}}
 \tilde{d}_{{\bf i}\alpha\sigma}^{\dag}
 \tilde{d}_{{\bf i}+{\bf a}\alpha\sigma}
 +U \sum_{{\bf i}}
 \tilde{\rho}_{{\bf i}\alpha\uparrow}\tilde{\rho}_{{\bf i}\alpha\downarrow}
 \nonumber\\
 &&
 - (\Delta/2) \sum_{\bf i}\cos\theta_{\bf i}\tilde{\rho}_{{\bf i}\alpha}.
\end{eqnarray}
Now we apply the second-order perturbation with respect to
electron hopping, and reach an effective Heisenberg spin model
\begin{equation}
\label{eq:ham-afm}
 \tilde{H}_{\rm AFM}=
 \sum_{{\bf i},{\bf a}}
 J_{{\bf i},{\bf i}+{\bf a}}
 ({\bf S}_{\bf i}\cdot{\bf S}_{{\bf i}+{\bf a}}-1/4)
 +E_{\rm s},
\end{equation}
where ${\bf S}_{\bf i}$ is a spin-1/2 operator at site ${\bf i}$,
the AFM exchange is given by
\begin{equation}
 J_{{\bf i},{\bf i}+{\bf a}}=
 4\vert \tilde{t}_{\alpha\alpha}^{{\bf i},{\bf i}+{\bf a}} \vert^2/U,
\end{equation}
and $E_{\rm s}$ denotes the energy shift
due to the level splitting, given by
\begin{equation}
 E_{\rm s}=-(\Delta/2)\sum_{\bf i}\cos\theta_{\bf i}.
\end{equation}

\begin{figure}[t]
\includegraphics[width=1.0\linewidth]{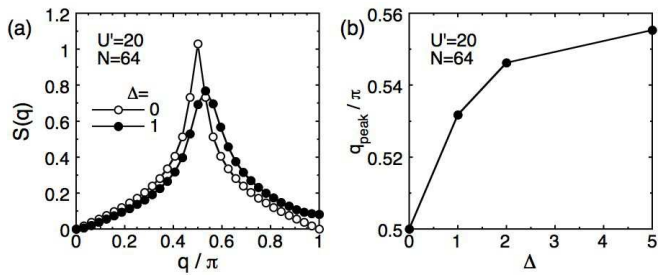}
\caption{
(a) Spin structure factor for $\Delta$=0 and 1 in the ground state.
(b) The $\Delta$ dependence of the peak position,
estimated from the DMRG results with $N$=64.
}
\end{figure}

It should be noted that
this effective model represents a zigzag spin chain
including nearest-neighbor exchange $J_{1}$
along the zigzag path
and next-nearest-neighbor exchange $J_{2}$
along the double chain.
To discuss magnetic properties of
the present $e_{\rm g}$-orbital model
based on the effective model,
let us here summarize the property of the zigzag spin chain,
which has been revealed by intensive numerical and analytical
investigations.~\cite{MG1969,Haldane1982,Tonegawa1987,Okamoto1992,White1996}
In the zigzag spin chain,
the ratio of competing exchanges $J_2/J_1$
is a controlling parameter for the degree of spin frustration.
At $J_2/J_1$=0,
the system is equivalent to a single chain,
which has a gapless spin-liquid ground state.
With the increase of $J_2/J_1$, the spin frustration becomes effective,
and the ground state is changed
from a critical spin-liquid phase to a gapped dimer phase
at $J_2/J_1$=0.24.~\cite{Okamoto1992}
The dimer phase continues up to the limit of $J_2/J_1$=$\infty$,
while the spin gap decreases exponentially,~\cite{White1996}
since the system turns to be independent two chains
for $J_2/J_1$=$\infty$.
On the other hand,
the spin correlation exhibits a commensurate AFM peak at $q$=$\pi$
for 0$<$$J_{2}/J_{1}$$<$1/2,
while the peak changes into an incommensurate one
for $J_{2}/J_{1}$$>$1/2,
and the peak position gradually moves to $q$=$\pi/2$
for infinite $J_{2}/J_{1}$.~\cite{Tonegawa1987,White1996}

In order to confirm that
magnetic properties are actually described by the zigzag spin chain,
let us discuss the behavior of the spin correlation.
In Fig.~4(a), we show our DMRG results of the spin structure factor
\begin{equation}
 S({\bf q})=
 \sum_{{\bf i},{\bf j}}
 \langle S_{\bf i}^{z}S_{\bf j}^{z} \rangle
 e^{i{\bf q}\cdot({\bf i}-{\bf j})}/N
\end{equation}
with
$S_{\bf i}^z$=$\sum_{\tau}
(\rho_{{\bf i}\tau\uparrow}$$-$$\rho_{{\bf i}\tau\downarrow})/2$.
Note that the results are depicted with a scalar wavenumber,
which is defined by regarding the zigzag path as a linear chain,
for comparison with the characteristics of the zigzag spin chain
described above.
For $\Delta$=0, we clearly observe a peak at $q$=$\pi/2$,
consistent with that of the zigzag spin chain with large $J_{2}/J_{1}$,
since $J_{2}/J_{1}$=$64^2$ for the 3$x^2$$-$$r^2$ FO arrangement.
Note that we evaluate $J_1$ and $J_2$ at the center of the chain,
by ignoring the site dependence in $J_{{\bf i},{\bf i}+{\bf a}}$.
The effect of the site dependence of $J_{{\bf i},{\bf i}+{\bf a}}$
on the spin gap will be discussed in the next section.
In Fig.~4(b),
we show the $\Delta$ dependence of the peak position of
the spin structure factor.
With increasing $\Delta$, the peak position changes
away from $q$=$\pi/2$ towards $q$=$\pi$,
as expected by analogy with the zigzag spin chain,
since $J_2/J_1$ approaches unity such as
$J_2/J_1$=1.61 for $\Delta$=1, $J_2/J_1$=1.24 for $\Delta$=2,
and $J_2/J_1$=1.08 for $\Delta$=5.
Note that in the zigzag spin chain,
the peak locates at $q$$\sim$$0.561\pi$ for $J_2/J_1$=1.~\cite{White1996}
In fact, the peak in the present $e_{\rm g}$-orbital model
approaches this position with increasing $\Delta$,
as shown in Fig.~4(b).
Thus, the behavior of the spin structure factor is well understood
by analogy with the zigzag spin chain.

\section{Spin Excited State}

Now we turn our attention to the spin excitation.
In order to discuss the spin gap, in general,
we consider the lowest-energy state among spin-triplet states.
One may expect that
such a triplet state is also described by the Heisenberg model,
but we should note that the change of the spin state influences
on the orbital state as well.
Such effect is already found in a two-site problem,
as shown in Fig.~5(a).
In the upper panel, we depict an AFM/FO configuration
in the spin-singlet ground state.
On the other hand, in the spin-triplet excited state,
an antiferro-orbital configuration turns to be favorable
to gain kinetic energy by electron hopping
due to the Pauli's exclusion principle,
as shown in the lower panel.

In order to clarify the characteristics of
the orbital rotation due to the spin-triplet excitation,
we investigate the spin-orbital correlation function,
defined by
\begin{equation}
 C_{\rm so}({\bf k},{\bf l};{\bf i},{\bf j})=
 \langle
 (3/4+{\bf S}_{\bf k}\cdot{\bf S}_{\bf l})
 T_{\bf i}^z T_{\bf j}^z
 \rangle,
\end{equation}
which represents the orbital correlation on the condition that
the spin state of bond $({\bf k},{\bf l})$ is triplet.
In Fig.~5(b), we show the spin-orbital correlation function
in the spin-triplet excited state.
Here, we set {\bf i}={\bf k} to consider the orbital rotation
around the spin-triplet bond.
We observe that staggered orbital correlation is enhanced
on the spin-triplet bond,
as was expected from the two-site problem,
while the staggered component rapidly decays
for long distances from the spin-triplet bond.
Namely, the orbital rotation occurs
only close to the spin-triplet bond.
Note that orbital correlations near the edges are found to increase
due to the deformation of the optimal orbital around the edges,
as shown below.

\begin{figure}[t]
\includegraphics[width=1.0\linewidth]{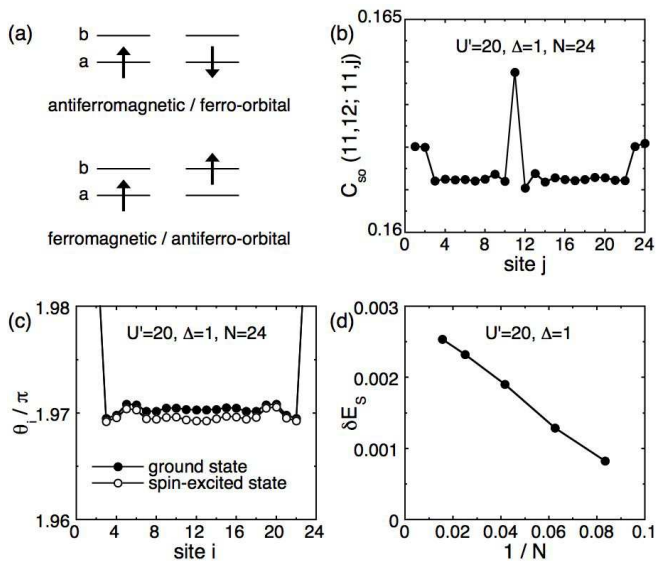}
\caption{
(a) Schematic view of spin-orbital configuration.
(b) Spin-orbital correlation function in the spin-excited state.
(c) The optimal set of $\{\theta_{\bf i}\}$
in the ground and spin-excited states.
(d) The system-size dependence of the difference in the energy shift
between ground and spin-excited states.
}
\end{figure}

Since the spin-triplet pair moves in the system
with carrying the orbital rotation,
it is expected that the averaged orbital structure should be different
between the spin-singlet ground state and the spin-triplet excited state.
In fact, in the excited state, a FO arrangement is also found to appear
and the system is regarded as a one-orbital system
in a similar way to the case of the ground state.
However, as shown in Fig.~5(c), we clearly observe that
optimal orbitals in the excited state are changed
from those in the ground state.
This behavior is understood from a viewpoint of
the adjustment of the optimal orbital.
Namely, the orbital state is reorganized so as to provide
the most appropriate orbital arrangement for the spin state.
Note that in the excited state,
the orbital shape around the edges becomes isotropic
in comparison with those in the bulk
similarly to the case of the ground state.

In order to ascertain this characteristic change of the orbital arrangement,
we evaluate the difference in the energy shift,
\begin{equation}
 \delta E_{\rm s}=E_{\rm s}^{(1)}-E_{\rm s}^{(0)},
\end{equation}
where $E_{\rm s}^{(0)}$ and $E_{\rm s}^{(1)}$ denote the values of
$E_{\rm s}$ in the ground and excited states, respectively.
In Fig.~5(d), we show the system-size dependence of $\delta E_{\rm s}$.
It is observed that $\delta E_{\rm s}$ grows and converges to a finite value
with increasing the system size,
suggesting that $\delta E_{\rm s}$ will contribute to the spin gap.
Concerning exchange interactions $\{J_{{\bf i},{\bf i}+{\bf a}}\}$,
we obtain the ratio at the center of the chain as
$J_{2}/J_{1}$=1.61 in the ground state,
while $J_{2}/J_{1}$=1.62 in the excited state for $\Delta$=1
from the DMRG results of $N$=40.
One may consider that the difference seems small,
but as we will see later,
even such small difference leads to the significant change
in the spin gap.
In short, the present system can be regarded as a one-orbital system even
in the excited state, but the variation of the orbital arrangement
appears in the changes of $E_{\rm s}$ and
$\{J_{{\bf i},{\bf i}+{\bf a}}\}$
in the effective Heisenberg model.
Thus, spin and orbital degrees of freedom dynamically
interact with each other, and the spin excitation involves
the changes of not only spin state itself, but also orbital state.

\begin{figure}[t]
\includegraphics[width=1.0\linewidth]{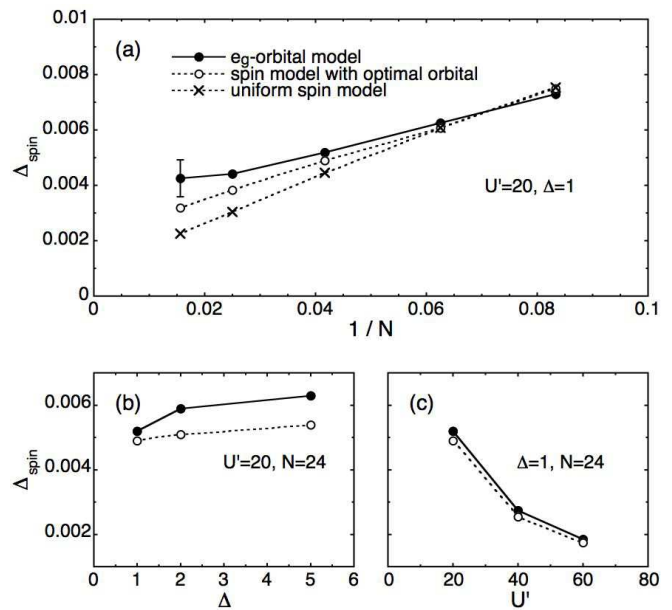}
\caption{
(a) The system-size dependence of spin gap at $\Delta$=1.
Errorbars for $N$=40 and 64 are estimated from the truncation error.
(b) Spin gap as a function of $\Delta$ for $U'$=20 and $N$=24.
(c) Spin gap as a function of $U'$ for $\Delta$=1 and $N$=24.
}
\end{figure}

Now let us discuss the behavior of the spin gap at $\Delta$=1.
From the estimation of $J_{2}/J_{1}$=1.61 in the ground state,
the ground state is expected to correspond to the gapped dimer phase
of the zigzag spin chain.~\cite{Tonegawa1987,Okamoto1992,White1996}
To clarify this point,
we investigate the system-size dependence of the spin gap
\begin{equation}
 \Delta_{\rm spin}=E(N/2+1,N/2-1)-E(N/2,N/2)
\end{equation}
up to $N$=64,
where $E(N_\uparrow,N_\downarrow)$ denotes the lowest energy
in the subspace with $N_\uparrow$ up- and
$N_\downarrow$ down-spin electrons.
As shown in Fig.~6(a), the spin gap seems to converge to
a finite value in the thermodynamic limit.
For comparison with the zigzag spin chain, we also evaluate
the spin gap for two types of AFM Heisenberg models.
First, we take a uniform model, assuming that AFM exchanges are
uniformly given by $J_{1}$ and $J_{2}$, which are estimated
at the center of the chain for the ground state.
In Fig.~6(a), such results are denoted by crosses.
Corresponding to the gapped dimer phase, we observe
a finite spin gap in the thermodynamic limit,
but significant deviation exists between the results of
the $e_{\rm g}$-orbital Hubbard model and
those of the uniform Heisenberg model.

It is emphasized here that such deviation indicates
the effect of the orbital-arranged background which should be
different between ground and excited states.
Then, we attempt to re-estimate the spin gap
by considering the following two points:
(i) The optimal orbital is not uniform in the system,
especially around the edges, as shown in Fig.~3(b).
(ii) The orbital arrangement is found to be different
between ground and excited states.
To take account of these points,
we deal faithfully with the effective Heisenberg model
Eq.~(\ref{eq:ham-afm}) with the optimal set of $\{\theta_{\bf i}\}$
for both ground and excited states.
The variation of $\{J_{{\bf i},{\bf i}+{\bf a}}\}$ causes
the change in the energy of the spin system.
There also occurs the difference in the energy shift
due to the level splitting, as shown in Fig.~5(d).
Note that we calculate the spin gap for the model
with site-dependent exchange interactions
by exploiting the DMRG algorithm
for random spin systems.~\cite{Hida1996}
The results are denoted by open circles in Fig.~6(a).
We find that the spin gap is appropriately
described by the effective Heisenberg model with the optimal orbital
rather than the simple uniform Heisenberg model.

Although some deviation still remains, it is improved
by increasing $m$ to keep more states in the renormalization step
for the $e_{\rm g}$-orbital Hubbard model.
Another aspect of the remaining deviation would be the effect of
charge and/or orbital fluctuations
due to the smallness of the Coulomb interaction
to arrive at the strong-coupling limit.
We note that larger energy is obtained for the effective Heisenberg model
than that for the $e_{\rm g}$-orbital Hubbard model
in both ground and excited states,
since the fluctuation effects are ignored.
However, such increase of the energy is found to be different
between ground and excited states.
As a result, the spin gap of the effective Heisenberg model
becomes small in comparison with the $e_{\rm g}$-orbital Hubbard model.
To see the fluctuation effects on the spin gap,
we show the $\Delta$ and $U'$ dependence of the spin gap
in Fig.~6(b) and 6(c), respectively.
With increasing $\Delta$,
$J_2/J_1$ approaches unity and the spin gap increases,
as expected by analogy with the zigzag spin chain,~\cite{White1996}
while the deviation remains significant
due to the effect of charge fluctuation.
On the other hand, as shown in Fig.~6(c),
the deviation is reduced with increasing $U'$,
indicating that the spin gap is well reproduced by the effective
Heisenberg model with the optimal orbital,
when the charge and orbital fluctuations are suppressed.
Thus, we conclude that active orbital degree of freedom influences
on the spin gap, suggesting that the spin gap should be considered
as a spin-orbital gap.

\section{Summary}

In this paper, we have discussed ground-state properties and spin gap
of the $e_{\rm g}$-orbital Hubbard model on the zigzag chain.
We have found that
the system is regarded as an effective spin system
on the orbital-arranged background due to the orbital arrangement.
At $\Delta$=0, two orbitals are degenerate,
but 3$x^2$$-$$r^2$ orbital is selectively occupied
so as to suppress the spin frustration due to the orbital anisotropy.
On the other hand, the orbital anisotropy is controlled by $\Delta$.
With increasing $\Delta$,
the lower 3$z^2$$-$$r^2$ orbital is favorably occupied
and the orbital anisotropy almost disappears,
leading to the revival of the spin frustration.
Note that the lower 3$z^2$$-$$r^2$ orbital is also occupied
in the atomic limit $U'$$\rightarrow$$\infty$ for $\Delta$$>$0.
Then, the system is corresponding to the gapped dimer phase
of the zigzag spin chain.
In actual compounds,
$\Delta$ and $U'$ could be controlled by hydrostatic or chemical pressure,
indicating the control of the degree of the spin frustration
and a possible realization of the gapped phase of the zigzag spin chain.

However, orbital degree of freedom is not completely quenched
due to the level splitting even at $\Delta$=1.
In fact, there appears a spatially non-uniform pattern of the optimal orbital
especially around the edges so as to adjust to the lattice inhomogeneity.
In addition, the orbital arrangement is flexibly deformed
according to the change of the spin state.
In order to discuss the spin-gap formation in multi-orbital systems,
it is quite important to consider the interplay
of spin and orbital degrees of freedom,
even when orbital degree of freedom is suppressed
due to the level splitting.
The spin gap is essentially determined by
the spin-orbital coupled excitation.
Thus, it is quite natural to consider the spin gap
as a spin-orbital gap.

\acknowledgments

We thank K. Kubo and K. Ueda for discussions.
T.H. has been supported by a Grant-in-Aid for Scientific Research
in Priority Area ``Skutterudites'' under the contract No.~18027016
from the Ministry of Education, Culture, Sports, Science, and
Technology of Japan.
He has been also supported by a Grant-in-Aid for
Scientific Research (C) under the contract No.~18540361
from Japan Society for the Promotion of Science.



\begin{thebibliography}{99}

\bibitem{QSS04}
{\it Proceedings of International Symposium on Quantum Spin Systems},
Prog. Theor. Phys. Suppl. {\bf 159} (2005).

\bibitem{Hase1993}
M. Hase, I. Terasaki, and K. Uchinokura,
Phys. Rev. Lett. {\bf 70}, 3651 (1993).

\bibitem{Masuda2000}
T. Masuda, I. Tsukada, K. Uchinokura,
Y. J. Wang, V. Kiryukhin, and R. J. Birgeneau,
Phys. Rev. B {\bf 61}, 4103 (2000).

\bibitem{Fukuyama1996}
H. Fukuyama, T. Tanimoto, and M. Saito,
J. Phys. Soc. Jpn. {\bf 65}, 1182 (1996).

\bibitem{Onishi2000}
H. Onishi and S. Miyashita,
J. Phys. Soc. Jpn. {\bf 69}, 2634 (2000).

\bibitem{Oosawa1999}
A. Oosawa, M. Ishii, and H. Tanaka,
J. Phys.: Condens. Matter {\bf 11}, 265 (1999).

\bibitem{Nikuni2000}
T. Nikuni, M. Oshikawa, A. Oosawa, and H. Tanaka,
Phys. Rev. Lett. {\bf 84}, 5868 (2000).

\bibitem{Ruegg2003}
Ch. R\"uegg, N. Gavadini, A. Furer, H.-U. G\"udel, K. Kr\"amer,
H. Mutka, A. Wildes, K. Habicht, and P. Vorderwisch,
Nature (London) {\bf 423}, 62 (2003).

\bibitem{MG1969}
C. K. Majumdar and D. K. Ghosh,
J. Math. Phys. {\bf 10}, 1388 (1969).

\bibitem{Haldane1982}
F. D. M. Haldane,
Phys. Rev. B {\bf 25}, 4925 (1982).

\bibitem{Tonegawa1987}
T. Tonegawa and I. Harada,
J. Phys. Soc. Jpn. {\bf 56}, 2153 (1987).

\bibitem{Okamoto1992}
K. Okamoto and K. Nomura,
Phys. Lett. A {\bf 169}, 433 (1992).

\bibitem{White1996}
S. R. White and I. Affleck,
Phys. Rev. B {\bf 54}, 9862 (1996).

\bibitem{Matsuda1995}
M. Matsuda and K. Katsumata,
J. Magn. Magn. Mater. {\bf 140-145}, 1671 (1995).

\bibitem{Matsuda1997}
M. Matsuda, K. Katsumata, K. M. Kojima, M. Larkin, G. M. Luke,
J. Merrin, B. Nachumi, Y. J. Uemura, H. Eisaki, N. Motoyama,
S. Uchida, and G. Shirane,
Phys. Rev. B {\bf 55}, R11953 (1997).

\bibitem{Kikuchi2000}
H. Kikuchi, H. Nagasawa, Y. Ajiro, T. Asano, and T. Goto,
Physica B {\bf 284-288}, 1631 (2000).

\bibitem{Hagiwara2001}
M. Hagiwara, Y. Narumi, K. Kindo, N. Maeshima, K. Okunishi,
T. Sakai, and M. Takahashi,
Physica B {\bf 294-295}, 83 (2001).

\bibitem{Maeshima2003}
N. Maeshima, M. Hagiwara, Y. Narumi, K. Kindo, T. C. Kobayashi,
and K. Okunishi,
J. Phys. Condens. Matter {\bf 15}, 3607 (2003).

\bibitem{Onishi2005a}
H. Onishi and T. Hotta,
Physica B {\bf 359-361}, 669 (2005).

\bibitem{Onishi2005b}
H. Onishi and T. Hotta,
Phys. Rev. B {\bf 71}, 180410(R) (2005).

\bibitem{Onishi2006}
H. Onishi and T. Hotta,
Physica B {\bf 378-380}, 589 (2006).

\bibitem{Brink1998}
J. van den Brink, W. Stekelenburg, D. I. Khomskii, and G. A. Sawatzky,
Phys. Rev. B {\bf 58}, 10276 (1998).

\bibitem{Bala2001}
J. Bala, A. M. Ole\'s, and G. A. Sawatzky,
Phys. Rev. B {\bf 63}, 134410 (2001).

\bibitem{Slater1954}
J. C. Slater and G. F. Koster,
Phys. Rev. {\bf 94}, 1498 (1954).

\bibitem{Hotta}
See, for instance,
T. Hotta, Rep. Prog. Phys. {\bf 69}, 2061 (2006).

\bibitem{Dagotto2001}
E. Dagotto, T. Hotta, and A. Moreo,
Phys. Rep. {\bf 344}, 1 (2001).

\bibitem{White1992}
S. R. White,
Phys. Rev. Lett. {\bf 69}, 2863 (1992).

\bibitem{Schllwock2005}
For review, see U. Schollw\"ock,
Rev. Mod. Phys. {\bf 77}, 259 (2005).

\bibitem{Hotta1998}
T. Hotta, Y. Takada, and H. Koizumi,
Int. J. Mod. Phys. B {\bf 12}, 3437 (1998).

\bibitem{Kugel1972}
K. I. Kugel and D. I. Khomskii,
Sov. Phys. JETP Lett. {\bf 15}, 446 (1972).

\bibitem{Kugel1973}
K. I. Kugel and D. I. Khomskii,
Sov. Phys. JETP {\bf 37}, 725 (1973).

\bibitem{complex-orbital}
Another type of complex orbital ordering has been discussed for manganites.
R. Maezono and N. Nagaosa,
Phys. Rev. B {\bf 62}, 11576 (2000);
J. van den Brink and D. Khomskii,
Phys. Rev. B {\bf 63}, 140416 (2001);
K. Kubo and D. S. Hirashima,
J. Phys. Soc. Jpn. {\bf 71}, 183 (2002).

\bibitem{Hotta2000}
T. Hotta, A. L. Malvezzi, and E. Dagotto,
Phys. Rev. B {\bf 62}, 9432 (2000).

\bibitem{Hida1996}
K. Hida, J. Phys. Soc. Jpn. {\bf 65}, 895 (1996).

\end{thebibliography}
\end{document}